# Positional distribution of human transcription factor binding sites.

Mark Koudritsky and Eytan Domany[*]

Department of Physics of Complex Systems, Weizmann Institute of Science, Rehovot

76100, Israel

[*]Corresponding author.
Email: eytan.domany@weizmann.ac.il
Tel: +972-8-9343964  Fax: +972-8-9344109

**Abstract**

We developed a method for estimating the positional distribution of transcription factor (TF) binding sites using ChIP-chip data, and applied it to recently published experiments on binding sites of nine TFs; OCT4, SOX2, NANOG, HNF1A, HNF4A, HNF6, FOXA2, USF1 and CREB1. The data were obtained from a genome-wide coverage of promoter regions from 8kb upstream of the Transcription Start Site (TSS) to 2kb downstream. The number of target genes of each TF ranges from few hundred to several thousand. We found that for each of the nine TFs the estimated binding site distribution is closely approximated by a mixture of two components: a narrow peak, localized within 300 base pairs upstream of the TSS, and a distribution of almost uniform density within the tested region. Using Gene Ontology and Enrichment analysis, we were able to associate (for each of the TFs studied) the target genes of both types of binding with known biological processes. Most GO terms were enriched either among the proximal targets or among those with a uniform distribution of binding sites. For example, the three stemness-related TFs have several hundred target genes that belong to "development" and "morphogenesis" whose binding sites belong to the uniform distribution.

# 1. Introduction

Elucidating the basic principles that underlie regulation of gene expression by transcription factors is a central challenge of the post-genomic era. Reliable experimental and computational identification of transcription factor binding motifs is an essential step towards this goal. In spite of major technological advances that generated rapidly improving high-throughput measurements of both gene expression [1] and transcription factor (TF) binding [2], and intense parallel bioinformatic efforts that produced a large variety of computational methods [3-6] aimed at identifying functionally important TF binding motifs, very basic questions remain unresolved. Perhaps one of the most pressing outstanding issues concerns the relative importance of proximal versus distal regulatory regions (with respect to the transcription start site (TSS)) in higher organisms.

While for prokaryotes the region in the close vicinity of the TSS is known [7] to play a central role in binding TFs that regulate gene expression, for eukaryotes the prevalent opinion is to the contrary; even though arguments supporting the special role of the proximal region have been presented for yeast [7] – it is believed that distal regulatory regions are most significant, especially for mammalians [8, 9]. Most recently several bionformatical studies have claimed that even in mammalians the proximal region dominates transcriptional regulation in general [10, 11] or for particular biological contexts [12]. There is no known estimate, however, of the relative abundance of distal as compared to proximal functional binding sites of transcription factors. There is no clear answer to simple questions such as the abundance of dual-action TFs, that under different conditions and in different pathways switch from proximal to distal regulatory binding. Conversely – do different genes, that belong to a particular biological function or pathway, exhibit the same positional bias in binding TFs that regulate their expression? The work presented here is an attempt to answer some of these questions by means of analysis of a large number of experimentally derived [13, 14] transcription factor binding sites.

To this end we developed a method for estimating the positional distribution of transcription factor binding sites on the basis of ChIP-chip data, and applied it to recently published experiments on binding sites of 9 transcription factors [13, 14], obtained from a genome-wide coverage of promoter regions from 8kb upstream of the TSS to 2kb downstream. Even though binding detected by ChIP-chip (in cell lines) is not synonymous to in-vivo functional binding that regulates transcriptional activity, knowing the positional distribution of binding sites does contain important, interesting and yet unexplored information.

The resulting estimated binding site distribution reveals an unexpected picture: it is closely approximated by a mixture of two components. One is a sharp peak, localized within 300 base pairs upstream of the TSS, and the second component is a distribution of almost uniform density within the tested region (-8kb to 2kb). These two components appear in all 9 transcription factors studied, but their relative weights do depend on the TF. Such a mixture of two distributions suggests that there might be two distinct groups of binding sites which differ in their biological function or in the mechanism by which their function is achieved. Indeed we found that the three TFs OCT4, SOX2 and NANOG, that constitute a control unit that governs the genetic program of embryonic stem cells [15], communicate with hundreds of genes involved in morphogenesis and development via uniformly distributed binding sites. On the other hand, the internal connections between these three TFs are of both kinds: the corresponding binding sites on the NANOG promoter are from the component proximal to the TSS, whereas on the other two promoters they are from the more distal uniform component. Further analysis and experiments are needed to elucidate other characteristics of the two kinds of binding sites and their possibly differing roles.

## 2. Materials and methods

**The data and platform:** We used ChIP-chip data from two studies. The first [13] aimed at mapping the binding sites of three transcription factors (TFs), NANOG, OCT4 and SOX2, known to play central roles in the maintenance of key properties of embryonic stem cells. The second study [14] concentrated on six TFs known to be expressed in the liver and believed to be critical for the biology of hepatocytes: HNF1A, HNF4A, HNF6, FOXA2, USF1 and CREB1. For 6 of the 9 TFs there is data from 2 biological replicas; for the remaining 3 – HNF6, USF1 and CREB1 only single replicas were available.

Both studies used human cells and the same custom designed DNA microarrays (code-named 10array) developed in the Young lab [16], containing 60-mer oligonucleotide probes. The probes cover regions that extend from 8kb upstream to 2kb downstream of the transcription start sites (TSS) of about 18,000 annotated human genes. On the average, there is approximately one probe every 280bp in the covered region. A full account of the technique can be found in supplementary material of [13] and [14] or on the web site accompanying those publications, http://jura.wi.mit.edu/young_public/hESregulation/Technology.html [17].

. Here we review only those components of the technique that are essential for understanding our analysis.

After immobilizing the proteins and fragmenting the DNA (into fragments of length of 550 bps on average), part of the resulting material is used for immunoprecipitation (IP), while the other part is reserved for control. The immunoprecipitation enriched DNA extract is labeled with red fluorescent dye while the control whole cell DNA extract is labeled with green. The whole cell extract (WCE) is assumed to contain any piece of the genome with equal probability (concentration), as opposed to the immunoprecipitated DNA extract that is significantly enriched by DNA fragments to which the TF of interest was bound. Both DNA extracts are applied to the microarray to allow competitive hybridization. The fluorescence intensity is then measured using red and green filters separately for each probe.

**Data analysis pipeline: identifying regions with bound TFs.** *I. Normalization and preprocessing* procedures (described in the Supplementary Information) were used, to assign each probe a score (referred to below as M-score) indicative of the probability of the presence of a binding site in its vicinity.

*II. Smoothed M-scores:* This score has unit variance and approximately normal distribution; hence M-score cutoffs can be interpreted in terms of probability. Since the spacing between adjacent probes was mostly within the resolution limit of chromatin immunoprecipitation (i.e. spacing was comparable to the DNA fragment lengths), a binding event on probe *i* was called on the basis of the M-scores of a triplet of consecutive probes and the value of their "triplet M-score",

$$M_i^{(3)} = (M_{i-1} + M_i + M_{i+1})/\sqrt{3}$$

Under the assumption of statistical independence of $M_i$ and $M_{i\pm1}$, the smoothed variable $M_i^{(3)}$ is also approximately normally distributed with unit variance (see Sec 1.3.3 of the Supplementary Information for verification of this). Using calls from a triplet of probes helps to filter out spurious signals from single isolated probes.

*III. Identifying bound triplets:* The filtering criterion (described in detail in Section 1.3 of the Supplementary Information) contains 4 different p-value-like cutoff parameters, $t_1$, $t_2$, $t_3$ and $t_n$, and an overall control parameter *(com)*.

A triplet centered on *i* was labeled as *bound* if it passed the following criteria:

(1) $M_i^{(3)} > (com) \cdot t_3$, AND

(2) either (2.1) $M_i > (com) \cdot t_2$ AND [$M_{i-1} > (com) \cdot t_2$ OR $M_{i+1} > (com) \cdot t_2$],

OR (2.2) $M_i > (com) \cdot t_1$ AND [$M_{i-1} > (com) \cdot t_n$ OR $M_{i+1} > (com) \cdot t_n$]

See supplementary Information for the rationale of these criteria, adopted from [13].

In order to avoid the arbitrariness often present when selecting a cutoff significance level, each of the four p-value-like cutoffs was initially assigned some reasonable value, adopted from [13], and these were then multiplied by the overall control parameter *(com)* (abbreviation for "cutoff multiplier"). The cutoff multiplier was varied from 0.1 to 500 and for each value the whole data analysis pipeline was run (lower multiplier value means stricter cutoff). For each TF we selected a "natural" value for the cutoff, as described in Results.

For each triplet of probes that passed the filters, the region between the two flanking probes (see Fig 1) was marked as a *bound region*. Overlapping bound regions were collapsed into a single bound region.

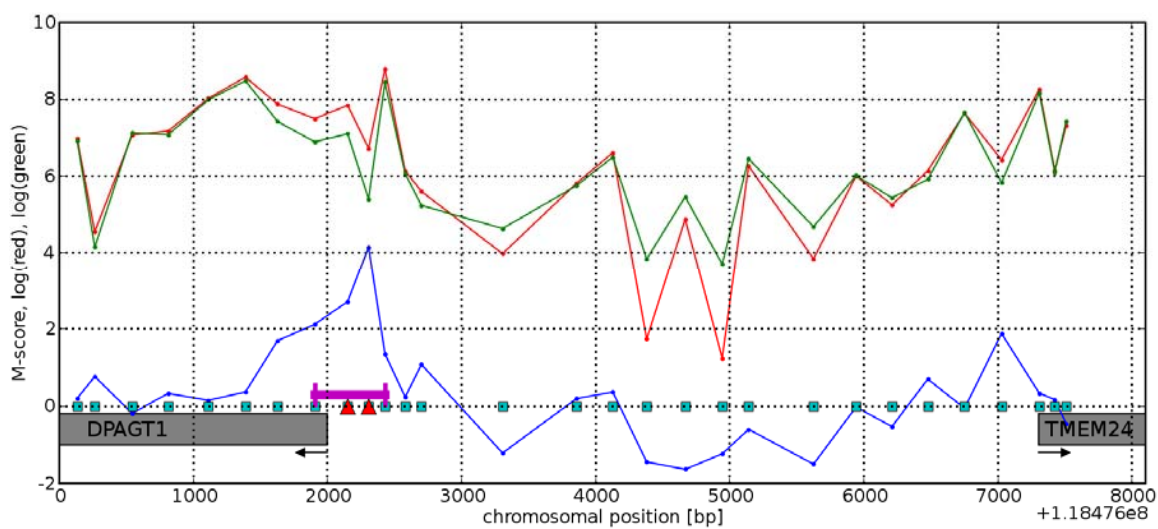

**Figure 1:** Example of a promoter region between TSSs of two genes: DPAGT1 and TMEM24, on chromosome 11. Microarray probes are depicted as squares on the *x* axis, red and green curves show log intensity of the red (IP) and green (WCE) channels from NANOG data, the blue curve is the resulting M-score. Probes detected as bound are marked with red triangles. The resulting bound region is marked with a magenta line. Arrows indicate direction of transcription.

The preprocessing steps described above resulted in lists of several hundred to several thousand bound regions for each TF. Each bound region is several hundred bps long (700 on average).

**Coverage plots.** In order to estimate the distribution of binding sites as a function of distance from transcription start site (TSS), promoters containing bound regions were aligned relative to the TSS nearest the bound region and a *coverage number* was calculated for each nucleotide location (defined with respect to its closest TSS). It is somewhat similar to a histogram, but since the bound regions have different lengths, a simple histogram could not be used. The coverage number of a particular position, at a given distance from the aligned TSSs, is the number of bound regions that contain the nucleotide at this position. That is, we count how many bound regions *cover* a point at distance *x* from the TSS, adding them up for all the genes tested. **Error! Reference source not found.** illustrates this concept. The genomic locations of the genes were taken from the RefSeq genes tablefrom UCSC genome browser, build hg17.

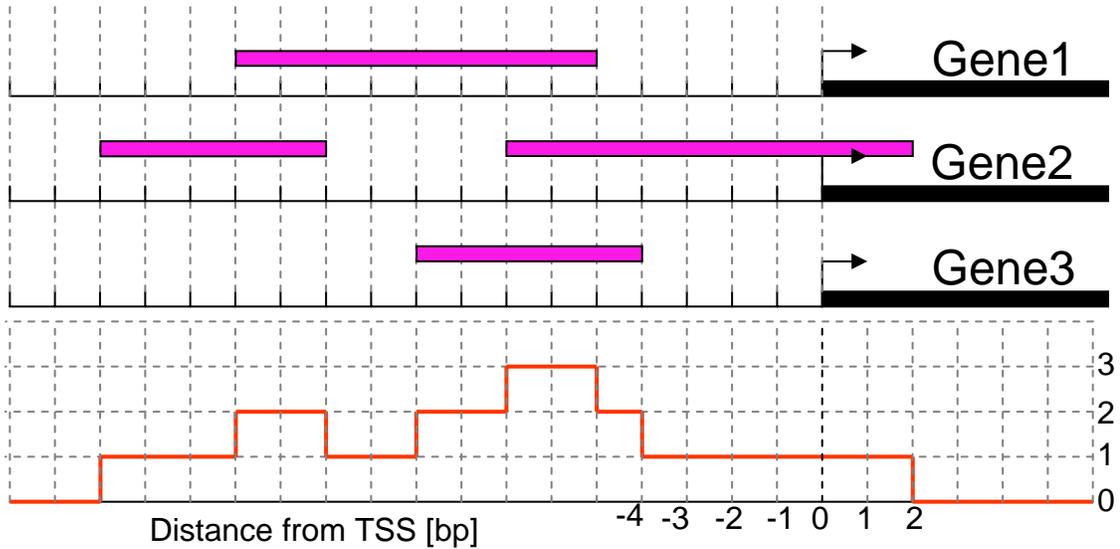

**Figure 2:** Illustration of the coverage number concept (not to scale). The red curve shows the coverage number of the hypothetical set of bound regions which are represented by magenta colored bars.

Coverage plots that were obtained from the experimental data for the nine TFs studied are presented in **Figure 3**. Note that in order to highlight the similarities between the different coverage plots, in each of these figures the coverage numbers are normalized by the area under the curves (the number of detected binding sites of the nine TFs varied between about 100 to 4000, see Table 1).

**Obtaining binding site distributions by deconvolution of coverage plots.** The genomic locations of the binding sites of a TF are *not known*; the aim of our analysis is to find, for each of the TFs, that distribution of binding sites (as a function of distance from the TSS), which provides the best fit for the corresponding (experimentally determined) coverage plot. The fit is obtained by a simulation of the entire process of TF binding events and their identification by the experiment.

Denote by $Q(x)$ the probability that the nucleotide at distance $x$ from the TSS belongs to a binding site of a given TF. We choose, for the TF that is studied, a particular $Q(x)$ from a family of distributions described below. A gene is randomly selected from the list of all possible targets, and a binding site is placed on its promoter at a location $x$ selected at random from $Q(x)$. Ten thousand binding sites are generated independently this way, each character-ized by a genomic address and a strength parameter $U$, which represents the binding affinity of the site. The value of $U$ is sampled at random from a shifted gamma distribution

$$p(U) = \frac{(U-s)^{k-1}}{\theta^k \Gamma(k)} \exp(-\frac{U-s}{\theta}) \qquad (1)$$

with parameters: shift $s=3$, shape $k=2$ and scale $\theta=3$, based on the model derived in [16]. The number of binding sites that were generated for a simulation (10,000) was chosen so that the resulting simulated coverage plot is not noisier than the experimental one. The precise value of the number of sites generated and the actual distribution of binding strengths had only a minimal effect on the results of our simulations.

Since the locations of all probes are known, we can calculate for each probe a simulated M-score, determined by its distance $d$ to the nearest of the 10000 simulated binding sites and by its strength parameter $U$:

$$\text{M-score} = f(d) \cdot U \qquad (2)$$

$$f(d) = \sum_{l=d}^{\infty} l \sum_{a=d}^{l} p(a) p(l-a) \qquad (3)$$

The influence function $f(d)$ used to calculate the M-score was adapted from the supplementary material of [18]; $p(a)$ is the probability that the DNA was cut at a distance $a$ from the nearest binding site. The distribution $P_L(l)$ of the DNA fragment lengths $l$ has been measured [18] and was approximated by a shifted gamma distribution [18]. This distribution is related to $p(a)$ by a convolution,

$$p_L(l) = \sum_{a=0}^{l} p(a) p(l-a) \qquad (4)$$

Since convolution of two identical shifted gamma distributions is again a gamma distribution (with twice the mean and shift), $p(a)$ was also taken to be a shifted gamma distribution. The following parameters were used for the simulation: shift $s=50$bp, shape parameter $k=2$ and scale parameter $\theta=60$. The simulated M-scores were then fed into the analysis pipeline described above as if they were derived from real raw data and a coverage number plot was generated. We performed simulations using binding site distributions $Q$ parametrized as described below and searched for a $Q$ that yielded a good approximation to the experimentally derived coverage plot. Since the simulation is computationally intensive (about 55sec on Intel

P4 2.4GHz 1GB RAM for a single run), any systematic fitting method would be difficult to implement and our forward-fit method has to be viewed as an approximate deconvolution.

**Family of binding site distribution functions tested.** The most prominent feature of the experimental coverage number graphs (see **Figure 3**) is the peak at around 150 bp upstream of the TSS. The structure away from this peak, namely the rapid decrease to zero at -8kb and +2kb (the edges of the genomic region covered by probes), is due to microarray design and is consistent with a uniform binding site distribution (see Results). Therefore we modeled $Q(x)$ as the sum of a uniform distribution and one or more Gaussians (up to 5, but usually one or two sufficed). The centers and widths of the Gaussians and the weights of all the components were the parameters that were varied to identify the distribution that gave best agreement with the experimental coverage plot.

**Gene Ontology (GO) enrichment analysis.** For a group of genes of interest $G$ (such as those targets of a TF whose binding is close to the TSS) we performed Gene Set Enrichment Analysis (GSEA) [19]. Admittedly over-representation of a specific GO category among the genes of $G$ does not *prove* regulation of these genes by the relevant TF, but co-regulation is a most plausible reason for the observed enrichment. To exclude the possibility that the over-representation of a GO group was due to a chance fluctuation in a random set of stray binding sites, the observed enrichment p-values were submitted to a stringent False Discovery Rate (FDR) analysis [20].

# 3. Results

**Coverage plots for the 9 tested Transcription Factors exhibit a prominent peak near the TSS.** The coverage plots obtained from the experimental data when the processing and analysis steps described above were implemented are shown in **Figure 3**. It is important to remember that the number of identified targets depends on the thresholds and cutoff parameters that were used; the dependence of the coverage plots on these parameters and the manner in which they were determined for each TF are discussed in detail below.

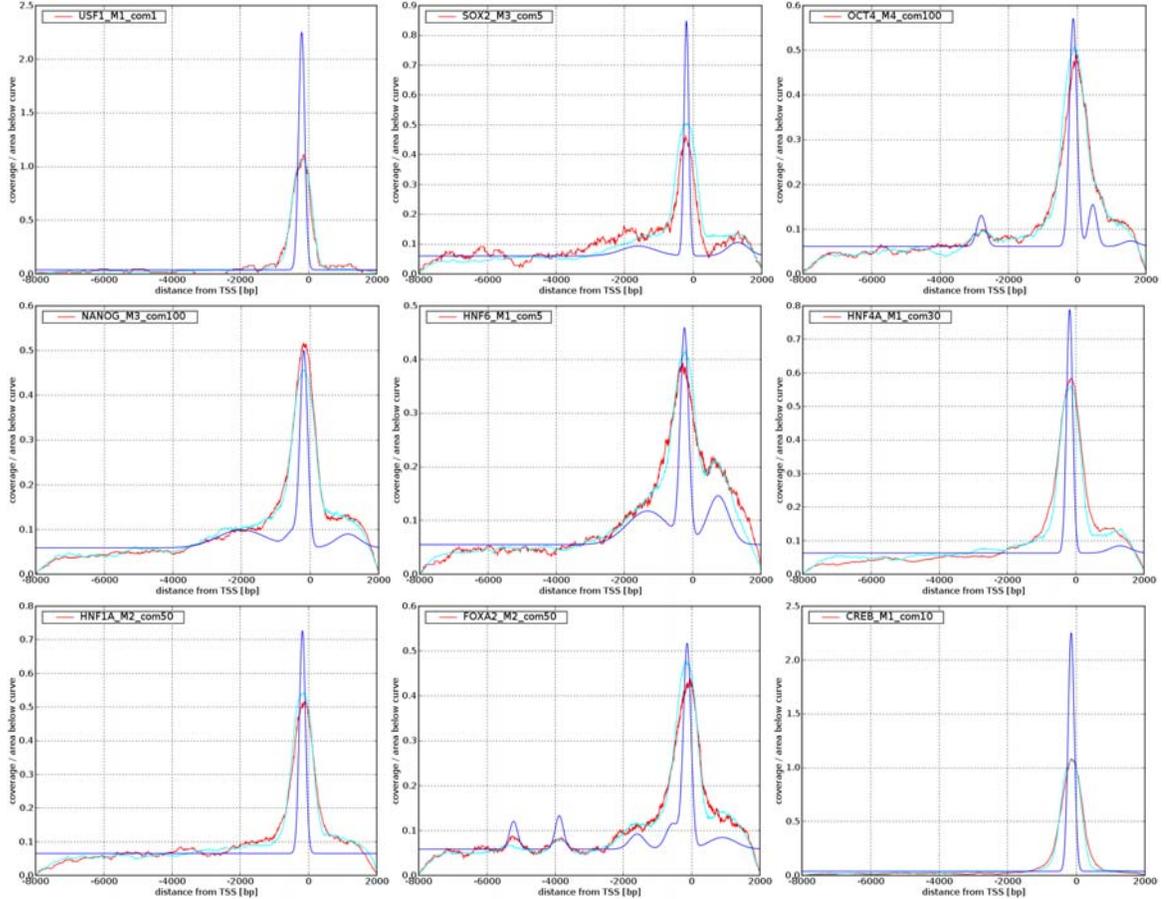

**Figure 3:** The fitted deconvolved binding site distributions (blue) and the corresponding simulated ones (cyan) compared with experimental (red) coverage number plots.

The coverage plots obtained from the experiments are in red; for 6 out of the 9 TFs the experiments were repeated twice, and coverage plots were prepared for each repeat. In 4 out of the 6 cases the two repeats are in good agreement with each other; in all 6 cases we chose to adopt the repeat with the higher peak (see Supplementary Figure S5). For each of the nine TFs studied the coverage plot exhibits a sharp peak close to the TSS. Th*is strong peak near the TSS is the most prominent feature of these plots*. As explained in Methods, these coverage plots were normalized, in order to emphasize that the peak near the TSS is shared by all 9 TFs studied.

**Binding site distributions obtained by fitting coverage plots.** In **Figure 3** we present for each of the 9 TFs studied the "optimal" binding site distribution $Q(x)$ that produced a good fit to the experimentally obtained coverage plots. The quality of the fit can be assessed from the same Figure. Good fits were obtained when a mixture of a uniform background distribution with one (or more) Gaussians was used. In several cases mixing a single sharp Gaussian, located near the TSS, sufficed; even when more than one was needed, the first one, located upstream close to the TSS, was the most prominent by far. The parameters of each of the binding site distributions, obtained for the nine TFs tested, are summarized in Table 1, together with numbers of bound regions (see Methods) and bound genes. A bound gene is defined here as any gene for which there is a bound probe in the interval from 10kb upstream to 3kb downstream of its TSS. Note that a single bound probe can give rise to more than one bound gene (sense and antisense).

| TF | *com* | $N_r$ | $N_g$ | $N'_g$ | $X_p$ | $D_f$ | $D_c$ | $W_u/W_p$ |
|---|---|---|---|---|---|---|---|---|
| NANOG_M3 | 100 | 2467 | 2394 | 1683 | -180 | 260 | 827 | 6.5 |
| OCT4_M4 | 100 | 1546 | 1536 | 623 | -120 | 260 | 914 | 5.7 |
| SOX2_M3 | 5 | 314 | 341 | 1271 | -200 | 165 | 624 | 5.7 |
| FOXA2_M2 | 50 | 1066 | 1126 | 890 | -130 | 260 | 862 | 6.1 |
| HNF1A_M2 | 50 | 1052 | 1097 | 1016 | -180 | 212 | 802 | 5.7 |
| HNF4A_M1 | 30 | 3889 | 3637 | 4519 | -180 | 212 | 860 | 5 |
| HNF6_M1 | 5 | 782 | 886 | 1306 | -240 | 283 | 1321 | 6.1 |
| CREB_M1 | 10 | 1008 | 1290 | 2197 | -150 | 212 | 708 | 1 |
| USF1_M1 | 1 | 111 | 151 | 1632 | -200 | 212 | 606 | 1 |

**Table 1:** Summary of the fitted distributions. *com* – CutOff Multiplier; it represents the p-value cutoffs selected for the TF as described later. $N_r$ - number of bound regions (bound regions are defined in Methods). $N_g$ - number of bound genes (a bound gene is defined here as any gene for which there is a bound probe within 10kb upstream to 3kb downstream of its TSS). $N'_g$ - number of bound genes as previously reported by [13, 14]. $X_p$ – peak position relative to TSS (in base pairs). $D_f$ – width in base pairs of the fitted peak at half max (2.36 σ), $D_c$ - width of the peak of the measured coverage plot at half max. $W_u/W_p$ - ratio of weights of the uniform component and of the peak in the distribution, it is also the ratio of the number of binding sites that are distributed uniformly and the number of binding sites that are localized within the peak.

The number of bound genes varies from about 150 for USF1 to 3600 for HNF4A. Note that for several TFs it differs considerably from the numbers reported previously by [13, 14]. These differences are due mainly to the different values of the p-value thresholds, selected as described later  The position of the peak is close to -200 bp for all TFs, and its width varies between 165 to 260. Using these numbers and by inspection of the deconvolved BS distribu-tions of Figure 3, we identified for each target gene the "proximal region" as the interval [-300,+300] bp (on both sides of the TSS), for all the TFs studied. If a particular region *R* was identified as proximal for gene *G1* and distal for *G2*, a binding event in *R* is counted as proximal binding to *G1* and distal to *G2*. Note that the width of the fitted distribution of binding sites is about 25% of the width obtained from the coverage plots; hence deconvolution of the coverage plots indeed sharpened significantly the resolution at which ChIP-chip data can be used to identify binding site positional bias.

The picture that emerges indicates that a TF may have two classes of binding sites, that probably differ in their biological function and the mechanism by which this function is achieved.

**Comparison of binding sites within the peak and outside using Gene Ontology (GO)**

We turned to look for such a difference of functions, using gene ontology [21] (GO) annotations of genes bound by the 9 TFs studied. For this end the genes bound by a particular TF were split into 2 disjoint groups – one contained genes that have a probe detected as bound, located within the gene's proximal region, and the other contains the rest of the TF's putative targets – i.e. genes with one or more bound probe, none of which lie within the gene's proximal region. The first group is assumed to include most of the genes that have a binding site on their promoter within the peak. Both groups were subjected separately to hypergeometric GO enrichment analysis (see section 2.5 of the Supplementary Information for details of the calculation of p-values and FDR correction), using only the "biological process" type of GO annotations. Results of this analysis are depicted graphically on Figure 4. Genes bound by HNF6, one of the six liver-associated TFs, were not enriched by any GO group. For the other TFs we do find association of biological processes with binding location; the GO groups are divided roughly into two subsets, (a) containing a variety of categories related to metabolism, RNA processing and splicing, cell cycle and more (see Supplementary Table S4 for a full list) and (b) mainly regulatory groups and developmental processes. The GO categories of subset (a) are enriched mainly by target genes of the liver-associated TFs and the binding sites are close to the TSS. The GO categories of type (b) are enriched mostly by genes that bind the three stemness related TFs, with binding sites far from the TSS. The finding that different GO terms are enriched in different groups supports the possibility of different functions associated with binding sites within and outside the peak. We list below a few selected observations.

For example, the group of genes with probes bound by OCT4, far from the TSS, is enriched with a GO term "organ morphogenesis" with an FDR corrected p-value of $4.2 \cdot 10^{-6}$, while the group of genes with probes bound by OCT4 *close* to the TSS is not enriched with the same term.

For some TFs such as USF1 and CREB1 there are enriched terms only in the group of genes with bound probes close to the TSS. The situation is reversed for OCT4. HNF4A has several GO terms enriched in both the far and the close groups. NANOG has some terms like mitosis enriched only in the close group, other terms, like morphogenesis, are enriched only in the far bound gene group, and yet others like RNA metabolism are enriched in both (see Table 2).

It is interesting to note that for genes bound by the stem cell TFs NANOG, OCT4 and SOX2, development related GO categories are enriched only among the genes with a binding site *far* from the TSS (see Table 2 and Supplementary Table S4).

**Figure 4:** Enrichment scores of about 100 GO terms among the genes bound by the studied TFs. Red color represent high enrichment. Each row is a GO term. The TFs are listed twice; left panel present the scores of enrichment among genes with proximal binding, while the right panel with distal – uniformly distributed binding. Notice the 2 clearly distinct groups of GO terms: one is predominantly enriched among the genes with proximal binding (the upper left corner) – those are mostly metabolism related GO terms and liver related TFs. The other group (bottom right corner) contains mostly development related GO terms enriched among genes with uniformly distributed binding sites of stem cell related TFs. Also note that NANOG is present in both of these groups.

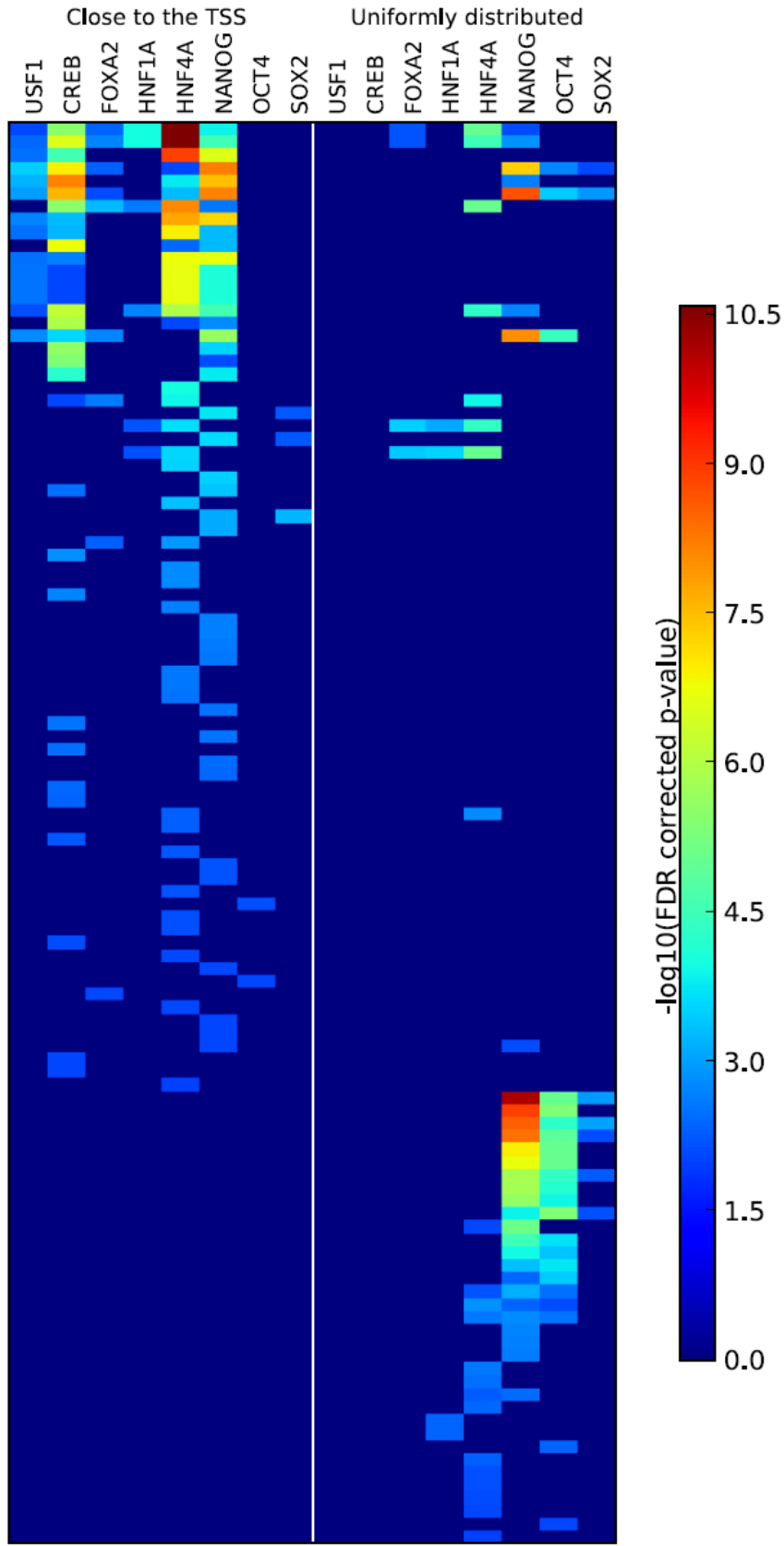

**Table 2:** GO categories enriched among the genes with a binding site of NANOG *far* from the TSS. Notice that there are several development related categories that are not enriched in the group of genes with binding site close to the TSS. See Supplementary Information for details about the calculation of p-values and FDR correction and a full list of enriched categories.

|  | FDR corrected P-values | |
|---|---|---|
| **GO category** | **Close to TSS** | **Far from TSS** |
| Development | 1.00E+00 | 7.02E-11 |
| Transcription | 6.37E-02 | 1.15E-09 |
| nucleobase, nucleoside, nucleotide and nucleic acid metabolism | 6.95E-09 | 2.01E-09 |
| anatomical structure development | 1.00E+00 | 2.88E-09 |
| organ development | 1.00E+00 | 4.35E-09 |
| RNA metabolism | 2.25E-06 | 9.47E-09 |
| biopolymer metabolism | 5.91E-09 | 4.83E-08 |
| RNA biosynthesis | 3.88E-02 | 1.16E-07 |
| transcription, DNA-dependent | 3.68E-02 | 1.94E-07 |
| Morphogenesis | 1.00E+00 | 1.42E-06 |
| regulation of nucleobase, nucleoside, nucleotide and nucleic acid metabolism | 4.06E-01 | 1.47E-06 |
| regulation of transcription | 4.13E-01 | 2.55E-06 |
| transcription from RNA polymerase II promoter | 1.52E-01 | 7.93E-06 |
| regulation of cellular metabolism | 4.16E-01 | 3.04E-05 |
| regulation of metabolism | 5.76E-01 | 1.17E-04 |

**Proximal and distal binding sites in transcriptional circuitry.**

It is interesting to analyze the different roles played by the two kinds of TF binding sites in the transcriptional circuits in which they participate. Figure 5 presents the connections *within* the three stem cell related TFs and their connections to the "external world" of transcriptional targets. Interestingly, NANOG itself is the target of exclusively proximal internal binding, whereas SOX2 and OCT4 have distal internal binding sites. As to external binding, the GO processes of development, morphogenesis, regulation of transcription and sensory organ development are controlled by binding sites from the distal, uniformly distributed class (of all three TFs).

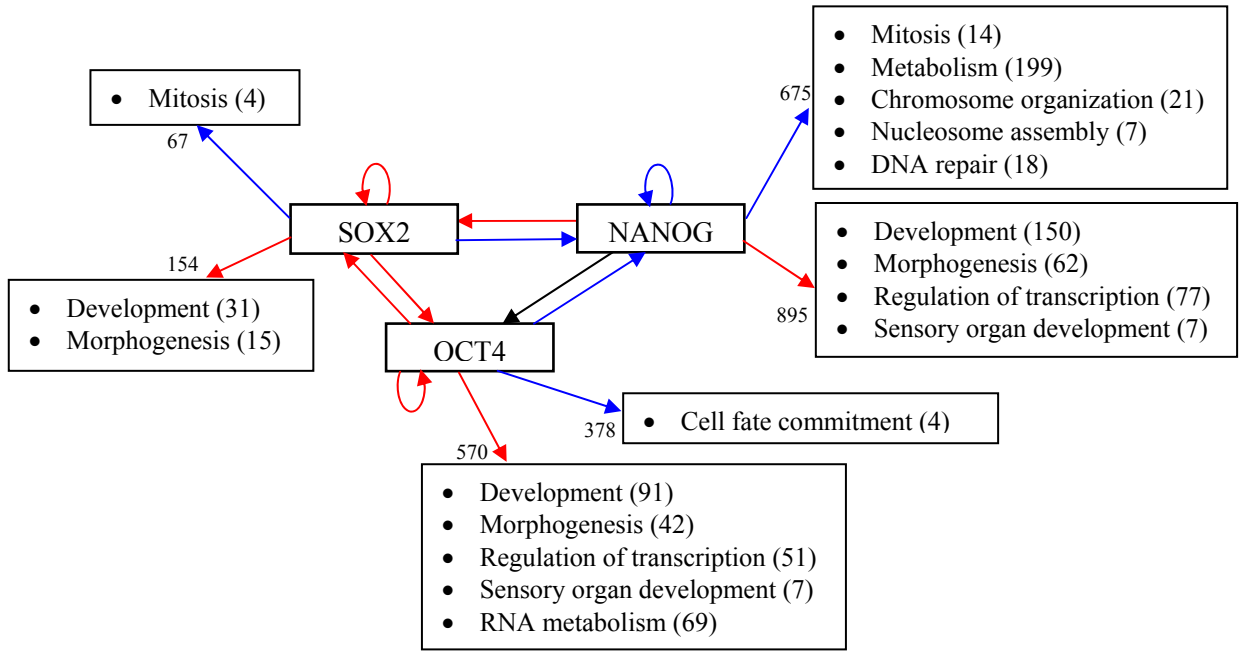

**Figure 5:** Schematic diagram of the stem cell circuit with some of the GO categories enriched among the genes bound by each TF. Blue arrows represent binding close to the TSS, red – distal, uniformly distributed. Black arrow means that binding is inferred from another source [15] and no information is available about the position. Numbers near the GO categories indicate the number of genes from the group in this category. Numbers on the arrows indicate the total number of genes in the group submitted to GO analysis (genes with multiple TSSs were omitted from this GO analysis). The information about the binding of OCT4 and SOX2 to the promoter of OCT4 was taken from [26] rather than from the ChIP-chip experiment since the microarray in the platform used does not cover properly the OCT4 promoter.

Another interesting observation concerns the circuit of liver related TFs: most of the internal interactions between the TFs in this group are either through binding close to the TSS or within the gene.

Having discussed in detail the main characteristics of the coverage plots and the fitted distributions obtained from their deconvolution, as well as the biological observations concerning the two kinds of binding sites we found, we turn to some technical details that must be addressed. First we rule out two possible trivial sources of the strong peak we found; next, present a purely computational test of the main result and describe the manner in which the p-value thresholds (that were used to identify binding events of the TFs studied) were set.

**Addressing several possible concerns about the analysis.**

We describe here several possible reasons and artifacts that could have misled us to reach the conclusion described above.

**The effect of probe density.** The first question to consider is whether the strong peak reflects nothing but the density of probes represented on the chip. Even though our simulations generate binding site distributions that fit the data (coverage plots) using the actual genomic locations of the probes placed on the chip, we wish to demonstrate here clearly that the distributions we found, and especially the sharp peak near the TSS, are not due to the probe distribution.

Clearly, if all probes are placed in a narrow region near the TSS, the coverage plot will have non-zero values in this region only. Indeed, since the microarray does not cover promoter regions outside the interval [-8kb, 2kb] from the TSS, we get zero coverage outside this region. Additionally, the probe density within the covered region is not uniform, as can be seen on **Error! Reference source not found.** (red curve): it is higher near the TSS. In order to understand how this probe density variation influences the coverage number, we performed a simulation of the measurement process starting with a hypothetical TF with a uniform distribution of binding sites as a function of distance from the TSS. The blue curve on **Error! Reference source not found.** is the average of the coverage number plots obtained from 100 such simulations. Comparison of this simulated curve with that generated from real data of HNF1A is presented on **Error! Reference source not found.**. Clearly, the peak of the real data is much sharper and narrower. Hence the prominent peaks observed for all 9 TFs cannot be attributed to the probe density variation, while the gradual decrease further away from the TSS most probably reflects just that.

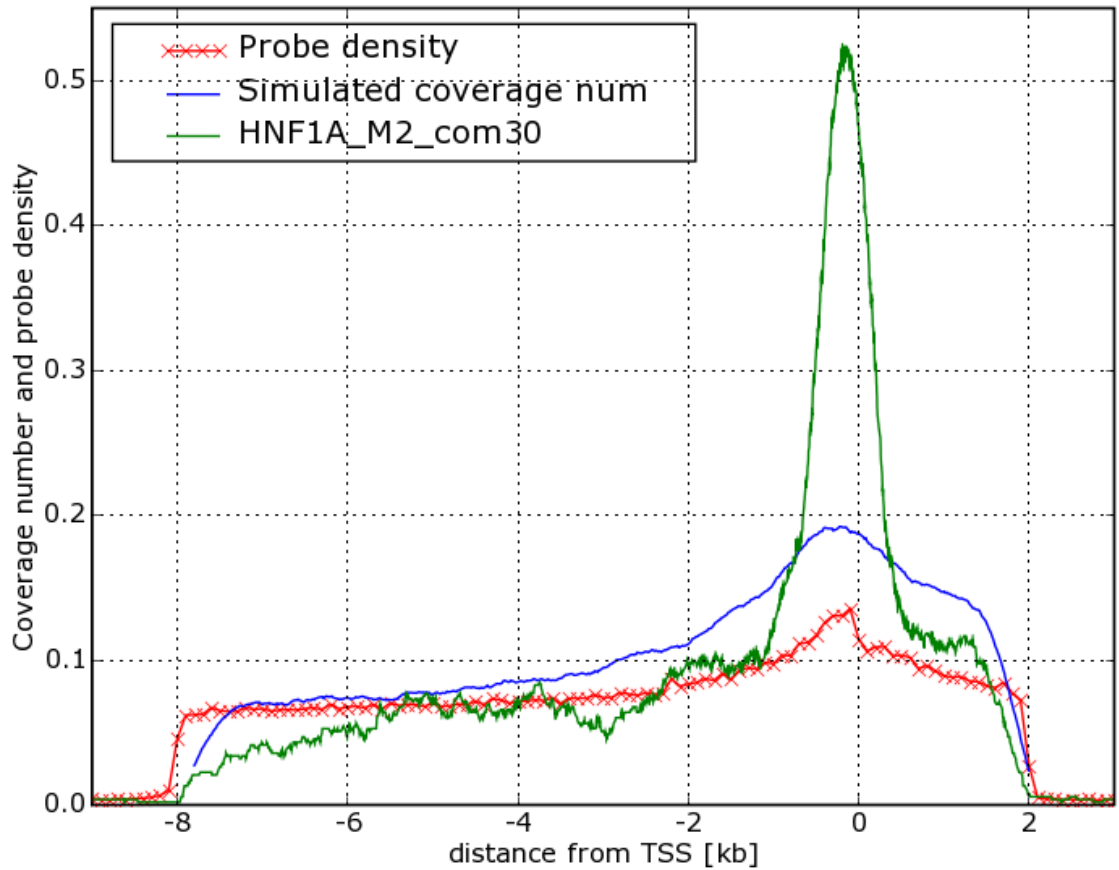

**Figure 6:** Comparison of coverage number plot for HNF1A with the coverage number plot obtained from a simulation that uses a uniform distribution of binding sites. All 3 curves were normalized to have the same area under the curve.

**The effect of GC-content variation.** A second possible artifact that could in principle produce such a peak is the non-uniform GC content of the promoters of human genes. As seen in Fig 7, the GC content increases from 45% far from the TSS to a peak value of about 65% near the TSS. Higher GC content means stronger binding of the DNA fragments to the corresponding probes and hence higher M-scores and, possibly, a higher density of detected binding sites. Since the bias introduced by the GC content is, on the average, similar for the IP and the WCE, we expect that taking the ratio of the intensities of the same probe in the two channels reduces significantly the effect of the GC variation on our measured signal. Indeed, as expected, the fluorescent intensity of the microarray probes in both channels is highly correlated with their GC content (correlation coefficient of about 0.7 in the data used here). On the other hand, correlation coefficients between the M-score and GC content are very low (order of 0.01) for most TFs, demonstrating that working with the M-score (which is basically a scaled log ratio of intensities in the two channels) successfully cancels the variation introduced by the GC content. As can be seen on **Error! Reference source not found.**, for NANOG the coverage number peak is about 150bp upstream from the TSS, while the GC content peak, as well as the peaks in intensity of the red and green channels, are about 70bp downstream. This significant shift in peak position serves as convincing evidence that the sharp peaks in coverage number plots are not an artifact caused by GC content variation.

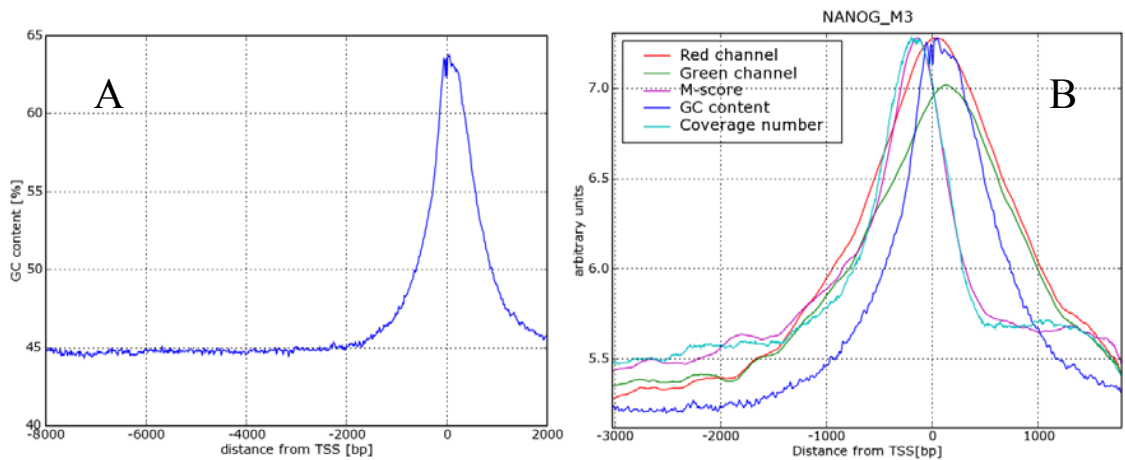

**Figure 7:**

**A)** GC content as a function of distance from TSS. Average over about 13,000 promoters, smoothed with a Gaussian kernel with $\sigma = 6$ nucleotides.

**B)** This figure shows the difference in location of the peaks of GC content and coverage number for NANOG averaged over all the promoters. Notice that the peaks of intensity of red and green channels coincide with the peak of GC content as expected. The peaks of coverage number and M-score, on the other hand, are more upstream the TSS providing convincing evidence that the sharp peaks in coverage number plots are not an artifact caused by GC content variation. The different curves were shifted and scaled vertically for convenient comparison; therefore the vertical axis has no meaningful units. The curves for M-score and the red and green channel intensities were obtained by linear interpolation between individual probes which was then averaged over all the promoters represented on the chip.

**A computational test.** To obtain further evidence for the fact that the peaks of the coverage plots and the resulting fits for binding site distribution are not due to some artifact of the method, we performed a purely computational test. We used the database underlying the "TFBS Conserved" track in the UCSC genome browser [23, 24], http://genome.ucsc.edu/cgi-bin/hgTrackUi?g=tfbsConsSites . It was generated using the TRANSFAC [25] collection of positional score matrices (PSSMs) representing the binding preferences of transcriptions factors. The database contains the locations and scores of transcription factor binding sites conserved in the human/mouse/rat alignment. In general, the number of conserved binding sites in this database is too small to construct meaningful positional histograms for most TFs, but for HNF1A, USF1, and CREB1 there was a large enough number. The resulting histograms are very similar to what we found from the ChIP-chip experiments (see Supplementary Figure S8). Since these coverage plots are derived in a purely computational way, they are not influenced by GC concentration in the same way as hybridization-based experiments.

**Are the identified binding sites functional?**

As stated in the Introduction, the binding events detected by ChIP-chip in cell lines may not necessarily correspond to functional binding, that actually regulates transcriptional activity, that takes place in vivo. Performing in vitro and in vivo experiments is the only way to establish beyond doubt the functionality of a binding site. Using in-silico bioinformatic methods to deduce functionality are at odds with the spirit and aims of this work, in which we tried to limit the analysis to experimentally derived binding events.

We did try to address specific concerns, in particular regarding a possible reasonable suspicion about functionality of the distal binding events. Over the 10Kbp long DNA strands scanned for binding one may (and will) have "stray" binding sites because of purely statistical reasons. We tried several tests, direct and indirect, to rule out the suspicion that our results reported above were based on such non-functional statistical binding events. As a sanity check of the assumed functionality of the distal binding sites, we investigated the promoters of a group of housekeeping genes (derived from [22]). Housekeeping genes are believed to be proximally regulated; we found that housekeeping genes had OCT4 and NANOG binding sites and, as expected, these had a much stronger tendency for proximal binding than the full genome-wide set of bound promoters. The weight ratio $W_u/W_p$ dropped from ~ 6 (see Table 1) to about 1.5 (see Supplementary Figure S14). Another (experiment-based) test is described below; by lowering the threshold for identification of a binding event from the data, we move from a regime where the identified binding events are dominated by strongly bound functional sites, to one where weaker stray statistical binding events constitute the majority. Since the origins of the two types of binding are very different, the number of detected binding sites should behave differently, as a function of the varying threshold, in the two regimes. Observation of such a difference (change of slope, apparent discontinuity, etc) is indicative of the fact that we indeed have two different types of binding sites, one of which is statistical and the other – most probably functional. We have demonstrated that for most of the studied TFs indeed such a crossover was observed for the distal binding sites (for which statistical binding occurs with high probability). These results are presented below and in Supplementary Figures S11. Since for 7 out of the 9 TFs studied the weight of the distal uniform distribution is about 6 times the weight of the proximal one (see Table 1), such a crossover induces a similar trend in the total number of binding sites (distal and proximal), as shown below.

**Selecting the p-value cutoffs for each TF**

As described in Methods, we used a single parameter to control the cutoff values of the four p-values that were used to decide whether a probe was considered bound or not by the TF. This *CutOff Multiplier* is referred to as *com* in the various figures and their legends. It was varied between 0.1 and 500 (lower values mean stricter cutoff, i.e. more rigorous filtering and smaller number of regions identified as bound). The numbers of bound regions and genes that were identified for each TF are reported in Table 1, which also contains (first column) the value of *com* used for each TF. Obviously the number of identified bound regions depends on the value of *com*, and we discuss here the manner in which we selected the values that were used. A related question concerns the extent to which the coverage plots, and in particular the sharp peak near the TSS, depend on *com*.

As described above, the general underlying assumption we make is that for low values of *com* we have very few false positives but many false negatives. As *com* increases, more binding events are identified, until at some point the resulting filter loses its meaning and the additional binding events are dominated by noise. Hence we are looking for a change of the behavior of either the number of binding events as a function of *com*, or of some other important property of the resulting coverage plots.

The observed behavior of coverage number plots as a function of cutoff can be divided into three different types. For five TFs the shape remained almost invariant as the cutoff multiplier increased, and deteriorated quickly beyond some "critical" value, above which the coverage plot resembled the one simulated for a hypothetical TF with uniform distribution of binding sites (Error! Reference source not found.). As shown in

Figure 8, HNF1A belongs to this type (the other four are FOXA2, HNF6, HNF4 and SOX2, see Supplementary Figure S9. As shown in Supplementary Figure S10, the total number of bound regions also exhibits a fairly sharp anomaly for these five TFs at the critical value of *com* (either a change of slope or apparent discontinuity). The critical value of *com* differs between transcription factors and may be different even between experimental replicas for the same TF (see Fig 6B).

A different behavior was exhibited by coverage number plots for NANOG (see Figures 6C and D) and OCT4 (see Supplementary Figure S9). For these TFs the peak value initially increases with *com* until a maximum is reached around *com* = 100, and then decreases.

The peaks of the coverage number plots of the remaining two TFs, USF1 and CREB1 decreased monotonically with *com* without any apparent discontinuities. It is interesting to note that these two TFs have the highest peaks, with coverage numbers nearly zero outside the peak (see **Figure 3**).

For the five TFs that exhibit the first type of behavior we selected *com* just below the critical value. The rationale is that for relatively stringent cutoffs we get coverage number plots that correspond to a relatively clean list of binding sites with few false positives. It can be assumed that while the relative coverage number plot does not change with loosening cutoff, the growing list of binding sites maintains a noise level similar to the initial one, until the critical value, beyond which many false binding sites enter the list and the noise level rises affecting the coverage number plot, and we choose *com* at a value before that happens.

The cutoffs for NANOG and OCT4 were selected to get the highest peak. Since the peak heights and number of bound genes for USF1 and CREB1 exhibited no obvious discontinuity (see Supplementary Figures S9 and S10) and therefore provided no hint for the selection of cutoffs, we selected rather conservative values of *com* (1 for USF1 and 10 for CREB1). This resulted in relatively small numbers of genes detected as bound by USF1 (and to some extent, by CREB1), compared to other TFs and to what was reported by [13, 14].

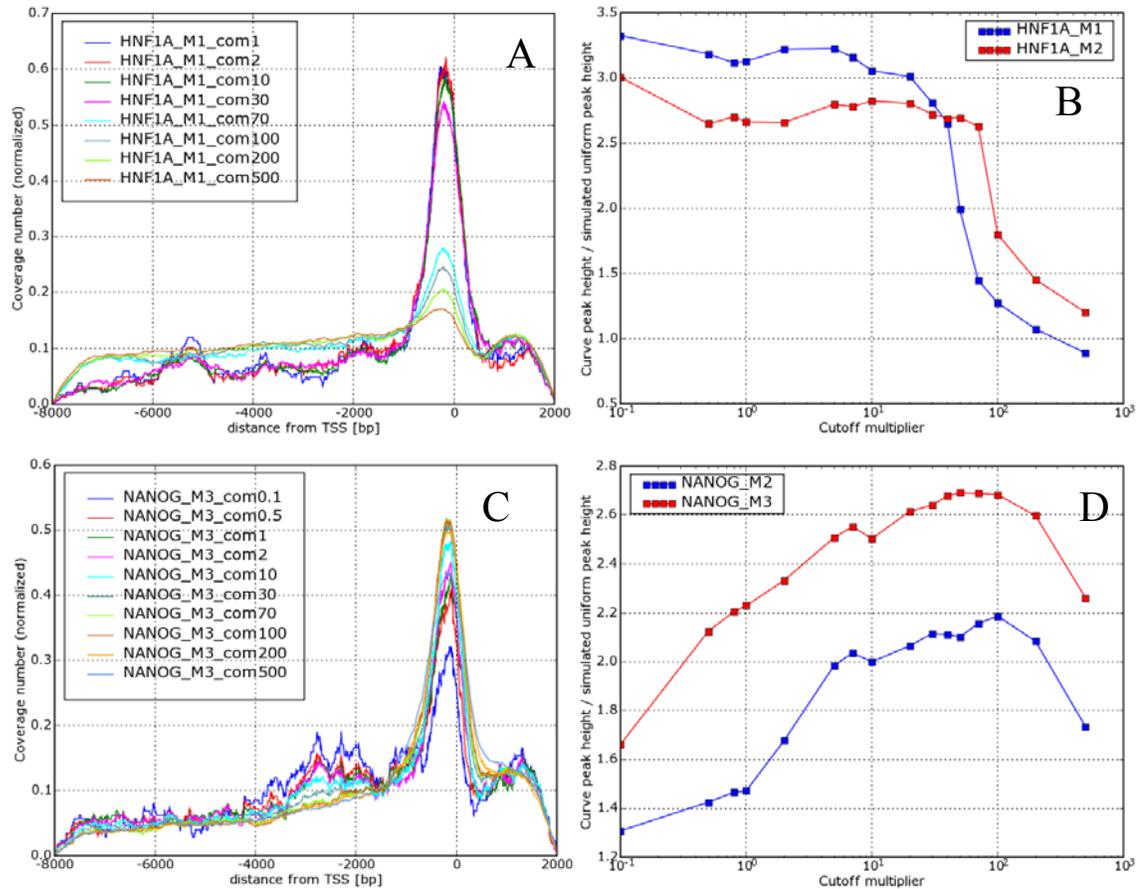

**Figure 8: A.** Coverage number plot and **B.** peak height for HNF1A; the same for NANOG (**C.** and **D**), for varying p-value cutoff multiplier (com). Note in A that for HNF1A the curve remains almost unchanged up to com70 while for NANOG it increases till *com* = 100 and diminishes again. The critical values of com, where the change of behavior occurs, depends on the TF and may vary also between replicate experiments of the same TF (B). The peak height plots are in units in which 1 is the height of the peak on a simulated coverage plot corresponding to a uniform distribution of binding sites.

# 3. Discussion

We used ChIP on chip data for nine transcription factors to pose and answer questions regarding the genome-wide distribution of binding sites with respect to the various genes' TSS. From the experiment we extracted *coverage plots*, from which we estimated the distribution of binding sites. This step was performed by using the experimental DNA fragment length distribution, the distribution of binding strengths and the actual addresses of the probes on the genome. The main result of this analysis is that the distribution of binding sites can be expressed as the sum of a very narrow peak close to the TSS, and a uniform background distribution.

We ascertained that our results are not due to various artifacts. For example, the effect of the non-uniform GC content (high near the TSS) on hybridization efficiency was assessed by a careful comparison with purely *in silico* results (available only for a subset of the TFs). The distortion caused by the non-uniform distribution of probes on the chip (denser near the TSS) was also taken into account. Thresholds of binding calls were set by a careful analysis of the dependence of the numbers of binding sites and the relative weights of the two components of the distribution on variation of the threshold. For most TFs we observed a fairly sharp change of behavior of these quantities, allowing us to identify the value of the threshold at which such changes set in, indicating a change in the strength of the contaminating noisy background signal. The number of target genes was found to be large. Even though most of the TFs studied were known to be hubs of transcriptional networks, the fact that the number of target genes of a TF is on the order of thousands (and that this seems to be the rule, rather than the exception!) seems to be fairly surprising.

Finally, we performed a functional analysis of the two types of genes: those that are regulated by binding sites proximal to the TSS and those whose binding sites belong to the uniform component of the binding site distribution. For the three TFs that regulate and govern embryonic stemness, we observed that the target genes associated with morphogenesis, development and regulation of transcription predominantly belong to the class with uniformly distributed regulatory binding sites. The molecular reasons behind this and the role this bias plays needs to be explored in the future.

**Funding**

This work was supported by the Ridgefield Foundation and by European Community (EC FP6) funding.

**Acknowledgments**

We thank Duncan Odom, Stuart Levine, Kenzie MacIsaac and Prof. Richard Young, for providing us with their raw data, for sharing their software and generously providing answers to numerous questions. Yuval Tabach, Tal Shay and Andrey Gubichev provided most helpful advice.